\documentclass[12pt,a4paper]{article}
\usepackage{graphicx}
\title{Multifractal analysis of the long-range correlations in the
cardiac dynamics of  Drosophila melanogaster}
\author{Nikolay K. Vitanov\,$^{\rm a,b}$ \footnote{corresponding 
author. e-mail: vitanov@imech.imbm.bas.bg, vitanov@mpipks-dresden.mpg.de},
Elka D. Yankulova \,$^{\rm c}$}
\date{
$^{\rm a}$ 
Max-Planck Institute
for the Physics of Complex Systems, N\"{o}thnitzerstr. 38, 01187 Dresden, Germany,
$^{\rm b}$
Institute of Mechanics and
Biomechanics of Bulgarian Academy of Sciences,
Akad. G. Bonchev Str., Block 4, 1113 Sofia, Bulgaria, $^{\rm c}$ Faculty of Biology, 'St. Kliment Ohridsky' University of Sofia,
8 Blvd. Dragan Tzankov, 1162, Sofia, Bulgaria, 
}
\begin{document}
\maketitle
\begin{abstract}
Time series of heartbeat activity of humans can exhibit long-range correlations.
In this paper we show that such kind of correlations can exist
for the heartbeat activity of much simpler species like
Drosophila melanogaster. By means of the method of multifractal detrended
fluctuation analysis (MFDFA) we calculate fractal spectra $f(\alpha)$ and $h(q)$
and investigate the correlation properties of heartbeat activity of Drosophila
with genetic hearth defects  for three consequent generations of species.
We observe that opposite to the case of humans  the time series of the
heartbeat activity of healtly Drosophila do not
have scaling properties. Time series from flies with genetic
defects can be long-range correllated  and can have multifractal properties.
The fractal heartbeat dynamics of Drosophila is transferred from generation to generation.
\end{abstract}
\section{Introduction}
The irregular and complex structure of the time series (ECG) of
human heartbeat dynamics is an object of considerable clinical
and research interest \cite{bassin}, \cite{ivanov1}, \cite{skinner1}.
This structure is connected not only to the external and internal perturbations but
also depends on the synergetic action of muscle and nervous systems which
influences the correlation properties of the time series.
In many simple systems the correlation function of the measured
time series usually decays
exponentially with the time. In complex systems
the correlations can decay with power law
and because no characteristic scale is associated with the power
law such systems are called scale-free. Their correlations are
called long-range because at large time scales the power law
function is always larger than the exponential function. Below we
are interested in the presence of long-range correlations  in
the time series for
heartbeat activity of Drosophila melanogaster - the classical
object of Genetics. Due to the short reproduction cycle of
Drosophila we can investigate the correlation properties of the heartbeat
dynamics for three consequent generations.
This allows us to study the relation
between genetic properties of Drosophila and correlation properties
of the time series of its heartbeat activity.
\par
The paper is organized as follows. In Sect. 2 we describe the
investigated system, recording of the time series and quantities
used for their analysis. The analysis of the obtained fractal spectra
is performed in Sect.3 . Some concluding remarcs are summarized
in the last section.
\section{System and methods}
We investigate  time series
of the heart activity (ECG) of Drosophila melanogaster
obtained from mutant flies and wild type controls provided by Bloomington
Drosophila Stock Center, U.S.A. We crossed male Dopa decarboxilase
(Ddc) mutant (FBgn 0000422 located in chromosome 2, locus 37C1)
and   female shibire (shi)  (FBgn 0003392 located in chromosome
1, locus 13F7-12). The Ddc mutants' heartbeat rate is about
60 \% of the normal one. Ddc codes for an enzyme necessary for the
synthesis of four neurotransmitters: norepinephrine, dopamine,
octopamine, serotonin, related to learning and memory. The shibire
(shi) mutants cause paralysis at high temperature. They code for
the protein dynamin, necessary for the endocytosis. Its damaging
at high temperature stops the transmission of the impulse through
the synapses, causes paralysis, and  eliminates the effect
of the neurotransmitters on the heart  \cite{johnson}. ECGs
were taken from three consequent generations of species. Drosophila
heartbeat was recorded optically  and digitalized. Optical ECG
records were taken at a stage P1 (white puparium) of a Drosophila
development when it is both immobile and transparent and the
dorsal vessel is easily viewed. The object was placed on a glass
slide in a drop of distilled water under a microscope
(magnification 350 x). Fluctuation in light intensity due to
movement of the dorsal vessel tissue was captured by photocells
fitted to the one eyepiece of the microscope. The captured
analogue signal was then digitized at 1 kHz sampling rate by data
acquisition card and LabVIEW data capturing software supplied by
National Instruments. 600000 data points  were taken for each
sample.
\par
The obtained time series are analysed by the multifractal
formalism which is widely used in mathematics, physics, and biology
\cite{bassin}, \cite{mandelbrot}, \cite{stanley1},
\cite{stanley2}, \cite{tel}. The investigation is
based on the spectrum $h(q)$ of the local Hurst exponent and
on the fractal spectrum $f(\alpha)$ \cite{peitgen,ott}.
Let us consider a set of points  which lies in an $N-$dimensional Cartesian space
covered by a grid of $N$-dimensional cubes of edge length $\epsilon$.
If for small $\epsilon$  we need $N^{*}(\epsilon)$ cubes to cover
our set we can define the spectrum of generalized dimensions
\begin{equation}\label{dqdim}
D_{q}=\frac{1}{1-q} \lim_{\epsilon \to 0} \frac{\ln I(q, \epsilon)}{
\ln (1/\epsilon)}, \hskip.5cm
I(q,\epsilon) = \sum_{k=1}^{N^{*}(\epsilon)} \mu_{k}^{q},
\end{equation}
where $q$ is a continuous index.
$\mu_{k}$ is the natural measure, i.e., it is a measure of the
frequency with which a typical orbit visits various cubes covering
the investigated attracting set of points for the limit case when
the length of the orbit goes to infinity (in addition the
frequences have to be the same for all initial conditions in the
basin of attraction of the attractor except for a set with
Lebesque measure $0$). Thus for $\mu_{k}$ we have
\begin{equation}\label{muk}
\mu_{k}=\lim_{T \to \infty} \frac{\xi (c_{k},{\bf x}_{0},T)}{T},
\end{equation}
where $\xi$ is the time the orbit originating from ${\bf x}_{0}$
spends in the cube $c_{k}$ in the time interval $0 \le t \le T$.
$D_{0}$ is called capacity of the set and it is not integer for
some sets. From (\ref{dqdim}) by means of the L`Hospital rule we can easily
obtain
\begin{equation}\label{d1}
D_{1}=\lim_{\epsilon \to 0} \frac{\sum_{k=1}^{N^{*}}(\epsilon) \mu_{i}
\ln \mu_{i}}{\ln \epsilon}
\end{equation}
$D_{1}$ is called also information dimension (as it measures
how the information is scaled with $\ln (1/\epsilon)$). In general $D_{0} \ge
D_{1} \ge D_{2} \ge \dots$. If $D_{q}$  varies with $q$ the measure, associated
with $D_{q}$ is called multifractal measure.
\par
Let a set $S$ be covered with a grid of
cubes of unit size $\epsilon$ and $\mu$ is the probability measure
on $S$ ($\mu(S)=1$). Let $\mu(c_{k})=\mu_{k}$ where $c_{k}$ denotes
again the $k-$th cube. We can assign a singularity measure
$\alpha_{k}$ to each cube
\begin{equation}\label{singmeas}
\mu_{k}= \epsilon^{\alpha_{k}}
\end{equation}
For small $\epsilon$ we can make continuous approximation for the
number of cubes for which $\alpha_{k}$ is between $\alpha$ and $\alpha+
d \alpha$, i.e., we can denote this number as
$\rho(\alpha) \epsilon^{-f(\alpha)} d \alpha$.
Substituting (\ref{singmeas}) in the relationship for $I(q,\epsilon)$
and after a transition from a sum over the cubes to an integration over
the $\alpha$ we obtain
\begin{eqnarray}\label{i1}
I(q,\epsilon)= \sum_{k=1}^{N^{*}(\epsilon)} \epsilon^{\alpha_{k} q} =
\int d \alpha^{*} \rho (\alpha^{*}) \epsilon^{-f(\alpha^{*})}
\epsilon^{q \alpha^{*}} =\nonumber \\
=\int d \alpha^{*} \rho (\alpha^{*}) \exp \left \{ [f(\alpha^{*})-
q \alpha^{*}] \ln (1/\epsilon) \right \}
\end{eqnarray}
For small $\epsilon$ \hskip.15cm $\ln (1/\epsilon)$ is large and
the main contribution to the above integral is from the
neighborhood of the maximum value of the $f(\alpha^{*})-q
\alpha^{*}$. Let $f(\alpha^{*})$ be smooth. The maximum is
located at $\alpha^{*}=\alpha(q)$ given by
\begin{equation}\label{max1}
\frac{d}{d \alpha^{*}} [f(\alpha^{*})-q \alpha^{*}] \mid_{\alpha^{*}=
\alpha(q)}=0 \to \frac{d f}{d \alpha^{*}}\mid_{\alpha^{*}=\alpha}=q
\end{equation}

\begin{equation}\label{max2}
\frac{d^{2}}{d (\alpha^{*})^{2}} [f(\alpha^{*})-q \alpha^{*}] \mid_{\alpha^{*}=
\alpha(q)}=0 \to \frac{d^{2} f}{d (\alpha^{*})^{2}}\mid_{\alpha^{*}=\alpha}=q
\end{equation}
Now we take the Taylor series representation of the
function $F(\alpha^{*},q)=f(\alpha^{*})-q\alpha^{*}$ around the point
$\alpha^{*}=\alpha (q)$ and  substitute it in (\ref{i1}). The result is
\begin{eqnarray}\label{i2}
I(q,\epsilon)=\exp \left\{ \right [f(\alpha(q))-q \alpha] \ln(1/\epsilon)\} \times
\nonumber \\
\times
\int d \alpha^{*} \rho (\alpha^{*}) \epsilon^{-(1/2)f^{ \prime \prime}
(\alpha(q)) (\alpha^{*}-\alpha(q))^{2}} \nonumber \\
\approx \exp \left\{ \right [f(\alpha(q))-q \alpha] \ln(1/\epsilon)\}
\end{eqnarray}
and a substitution of relationship (\ref{i2}) in (\ref{dqdim}) leads to
\begin{equation}\label{dqdim2}
D_{q}=\frac{1}{q-1} \left[ q \alpha(q) - f(\alpha(q)) \right]
\end{equation}
Using (\ref{max1}) we obtain
\begin{equation}\label{dqx}
\frac{d}{d q} \left[ (q-1) D_{q} \right] = \alpha (q) = \frac{d \tau}{d q}
\end{equation}
Then
\begin{equation}\label{tauq}
\tau(q) = (q-1) D_{q} \to D_{q}= \frac{\tau (q)}{q-1}
\end{equation}
From (\ref{dqdim2})
\begin{equation}\label{f2}
f(\alpha(q)) = q \frac{d \tau}{d q} -(q-1) D_{q} = q \frac{d \tau}{d q} -
\tau (q)
\end{equation}
For each $q$ from (\ref{tauq}) and (\ref{f2})
we can obtain $\alpha (q)$
and $f(\alpha)$ thus parametrically specifying the function $f(\alpha)$.
And $\alpha$ can be connected to  the local Hurst exponent by means of the relationships
\begin{equation}\label{alpha1}
\alpha = h(q) + q \frac{d h}{dq}, \hskip.5cm
f(\alpha) = q [\alpha - h(q)] +1
\end{equation}
Thus obtaining the $h(q)$ spectrum we can obtain also $\alpha$ and
$f(\alpha)$ spectra by means of (\ref{alpha1}).
\par
For calculation of $h$ from the heartbeat time series we can
use the method of  multifractal detrended fluctuation analysis
(MFDFA) or the more complex  wavelet transform modulus maxima
method (WTMM), initially developed for investigation of quasi-singularities
of turbulent signals (for applications of this method see \cite{muzy, arneodo},
\cite{arneodo96}, \cite{arneodo99}, \cite{arneodo03},
\cite{dimitrova}). In this paper we shall use the MFDFA method which
realization is as follows  \cite{kantelhardt}.
First of all we have to calculate  the profile function $Y_{i}$. For this we
calculate the mean $\langle x \rangle$ of the investigated time series $\{x_{k}\}$ and
use it to obtain the profile function
\begin{equation}\label{profile}
Y_{i}=\sum_{k=1}^{i} (x_{k} - \langle x \rangle), \hskip.5cm i=1,2,\dots,N.
\end{equation}
The following step is to divide the time series into segments and  to
calculate the variation for each segment.
The division is into $N_{s}=$int$(N/s)$ segments and because the obtained segments would
not include some data at the end of the investigated time series,  additional $N_{s}$ segments
are added, which start from the last value of the sequence in the direction to the first
value of sequence.
\par
In order to calculate the variation  we have to calculate the
local trend (the fitting polynomial $y_{\nu}(i)$ for each
segment of length $s$,
where $s$ is between an appropriate minimum and  maximum value).
Then the variations are defined as
\begin{equation}
F^{2} (\nu,s) = \frac{1}{s} \sum_{i=1}^{s} \left \{ Y[(\nu-1) s+i] -y_{\nu}(i)
\right \}^{2}
\end{equation}
for the first $N_{s}$ segments  and
\begin{equation}
F^{2} (\nu,s) = \frac{1}{s} \sum_{i=1}^{s} \left \{ Y[N-(\nu-N) s +i] -
y_{\nu}(i) \right \}^{2}
\end{equation}
for the second $N_{s}$ segments.
Finally we construct the $q$-th order fluctuation function
\begin{equation}
F_{q}(s) =  \{ [1/(2N_{s} )] \sum_{\nu=1}^{2N_{s}} [ F^{2} (\nu,s)]^{q/2}
\}^{1/q}.
\end{equation}
The scaling properties of $F_{q}(s)$ determine the kind of fractal characteristics
of the time series. For monofractal time series $F_{q}(s)$  scales as $s$ of constant
power $h$ for each $q$. For sequences of random numbers this constant $h$
has the value $1/2$. Even in presence
of local correlations extending up to a characteristic range $s^{*}$ the
exponent $h=1/2$ would be unchanged when $s>>s^{*}$. If the correlations do
not have characteristic lengths the exponent $h$ would be different from $1/2$.
\par
The procedure described above is appopriate for determination of
positive Hurst exponents which are not very close to zero.
For close to zero or negative exponents we have to add a step
after the calculation of the profile function  namely to calculate
the profile function of the profile function $Y$
\begin{equation}\label{profile2}
Y^{*}_{i}=\sum_{k=1}^{i}[Y(k)-\langle Y \rangle ]
\end{equation}
and the function $Y^{*}_{i}$ should be used further in the MFDFA procedure.
The result is that if there is a scaling in the fluctuation function this
scaling is connected to the Hurst exponent as
\begin{equation}\label{hoeld2}
F^{*}_{q}(s) \propto s^{h(q)+1}
\end{equation}
In our investigation below we use MFDFA(1) i.e. the local trend for each segment
is approximated by a straight line.
\begin{figure}[t]
\centerline{\includegraphics[angle=270,width=12cm]{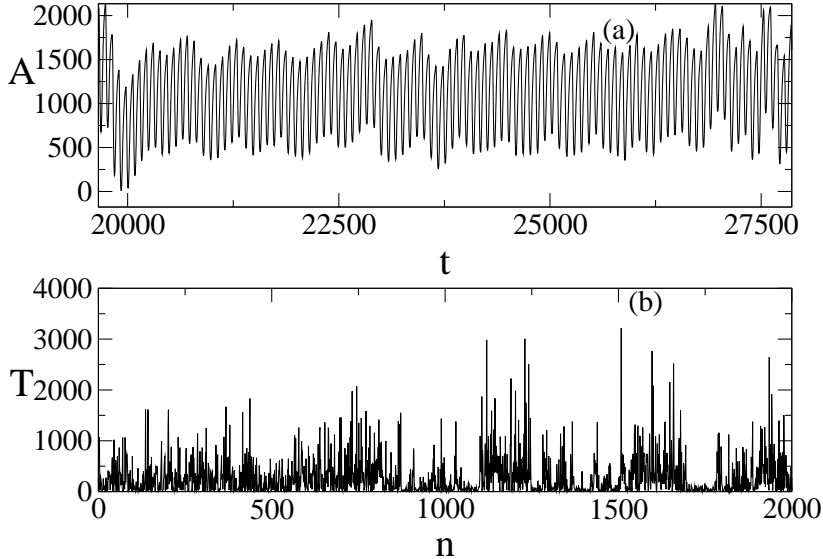}}
\caption{Panel (a): Typical time series of the heart activity of
Drosophila melanogaster. The unit for time is $0.001$ s. Panel
(b): Interbeat intervals for the time series of heart activity of
Drosophila. As we can see the time series of the heart rate
fluctuate irregularly from beat to beat. }\label{fig:01}
\end{figure}
\section{Results}
A part of typical time series for the heart activity of Drosophila melanogaster is
presented in panel (a) of Fig. 1. From these time series we can construct time series for the
interbeat intervals (presented in panel (b) of Fig. 1). Such
time series are widely studied for humans \cite{peng1},
\cite{ivanov2} because they can be easily measured in a noninvasive
way and may have diagnostic and prognostic value.
The interbeat time series of human heartbeat dynamics has (i) monofractal properties (constant $h$) for humans
with heart diseases and (ii) multifractal properties (nonconstant $h$)
for time series from healtly humans.
As we shall see this is not the case for Drosophila.
\begin{figure}[t]
\centerline{\includegraphics[angle=270,width=12cm]{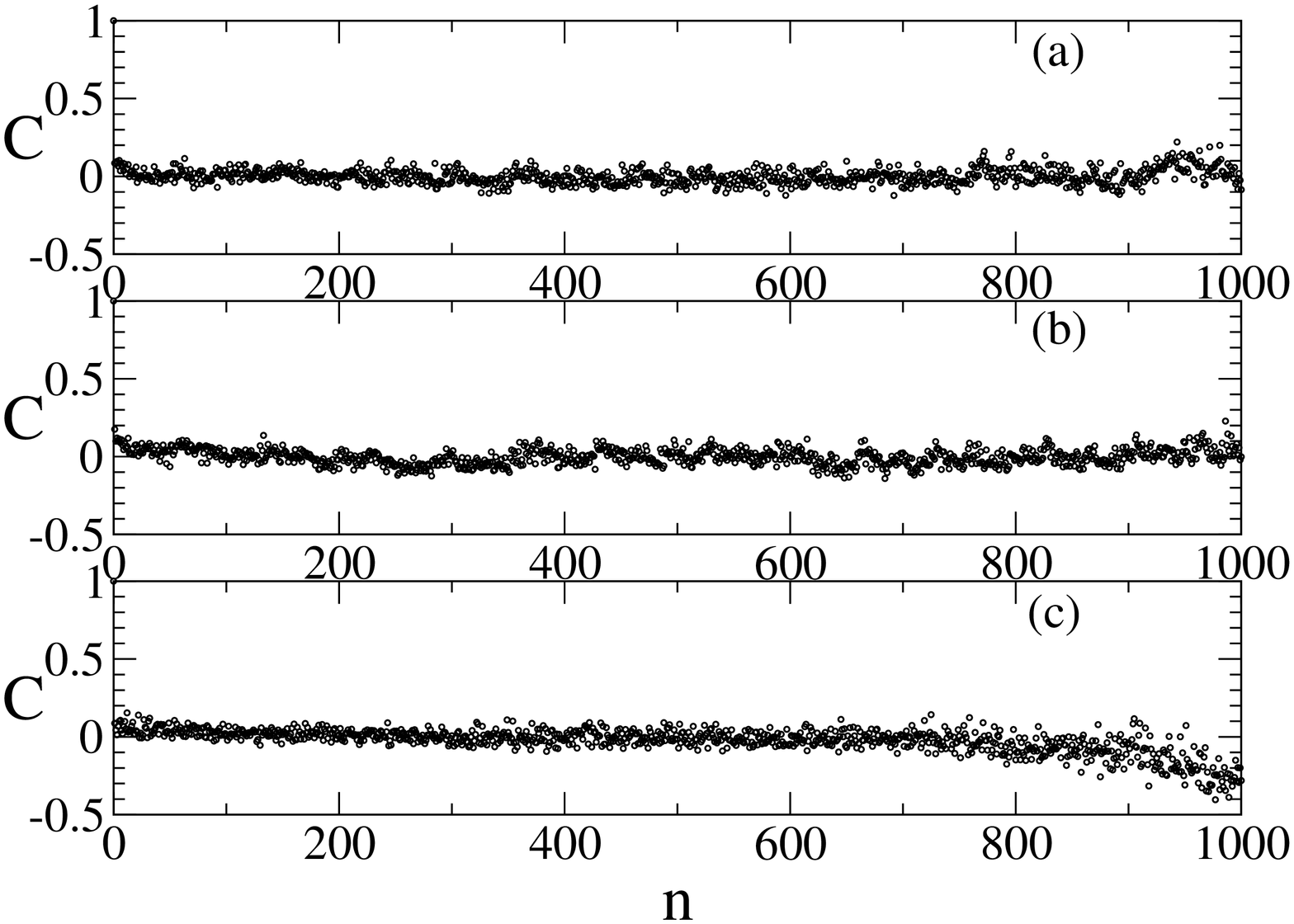}}
\caption{Autocorrelation function $C(n)$ for the time series of the heart activity of
Drosophila. Panel (a) : autocorrelation for a
healtly animal. Panels (b) and (c): autocorrelations for the two parents: female (panel (b))
and male (panel (c)).
}\label{fig:02}
\end{figure}
\par
The autocorrelation functions for a healtly control fly and for the parents
with heart defects are shown in Fig. 2. In all three panels we observe that a
significant degree of correlation exists even for large values of $n$.
In addition in panel (c)
we observe systematic decrease of the autocorrelation function and transition from
predominantly correlated behavior for small $n$ to predominantly anticorrelated behavior
for large $n$. Thus the dynamical consequences of the different genetic heart defects of
Drosophila are clearly visible.
\begin{figure}[t]
\centerline{\includegraphics[angle=270,width=7cm]{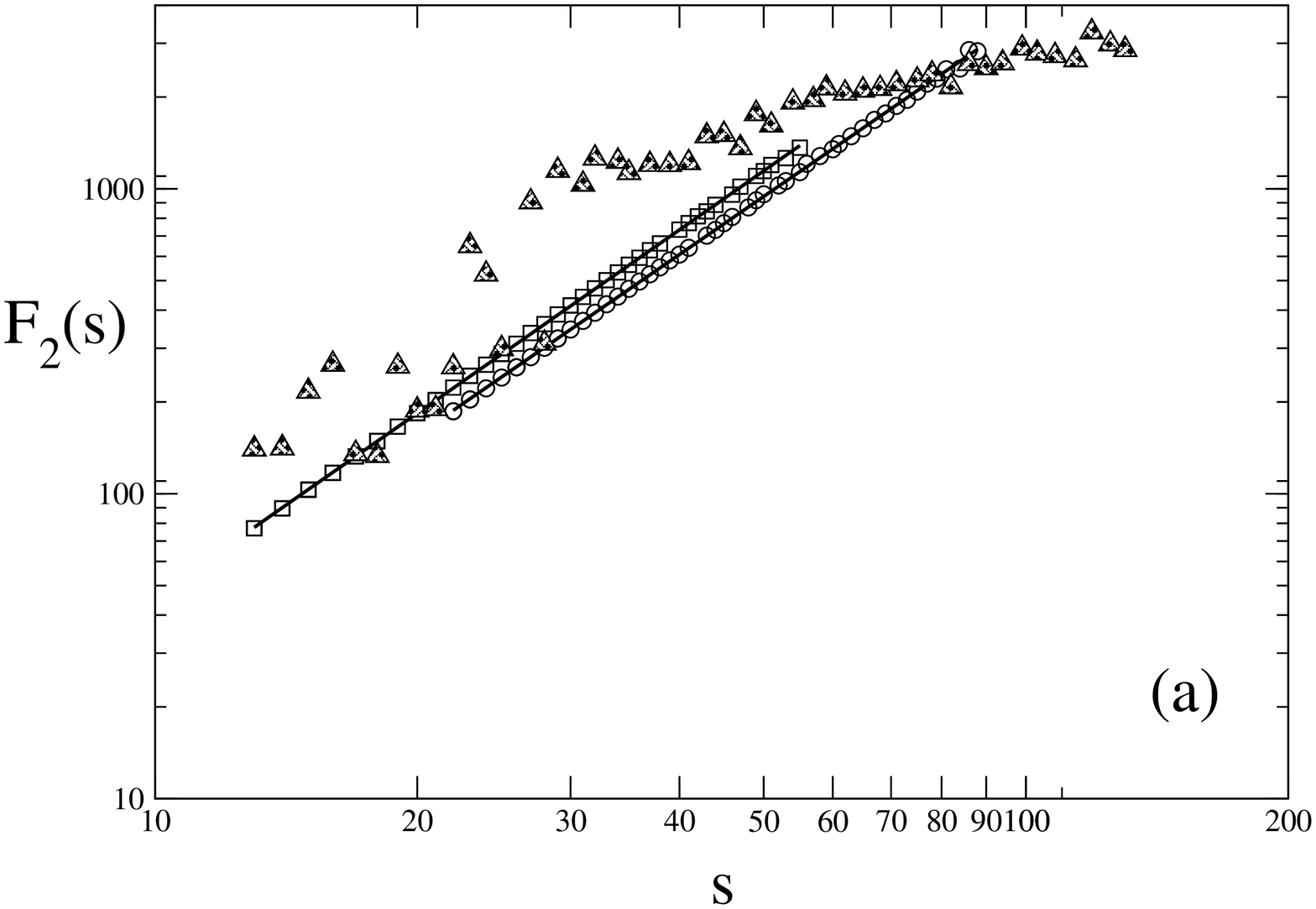}
\includegraphics[angle=270,width=7cm]{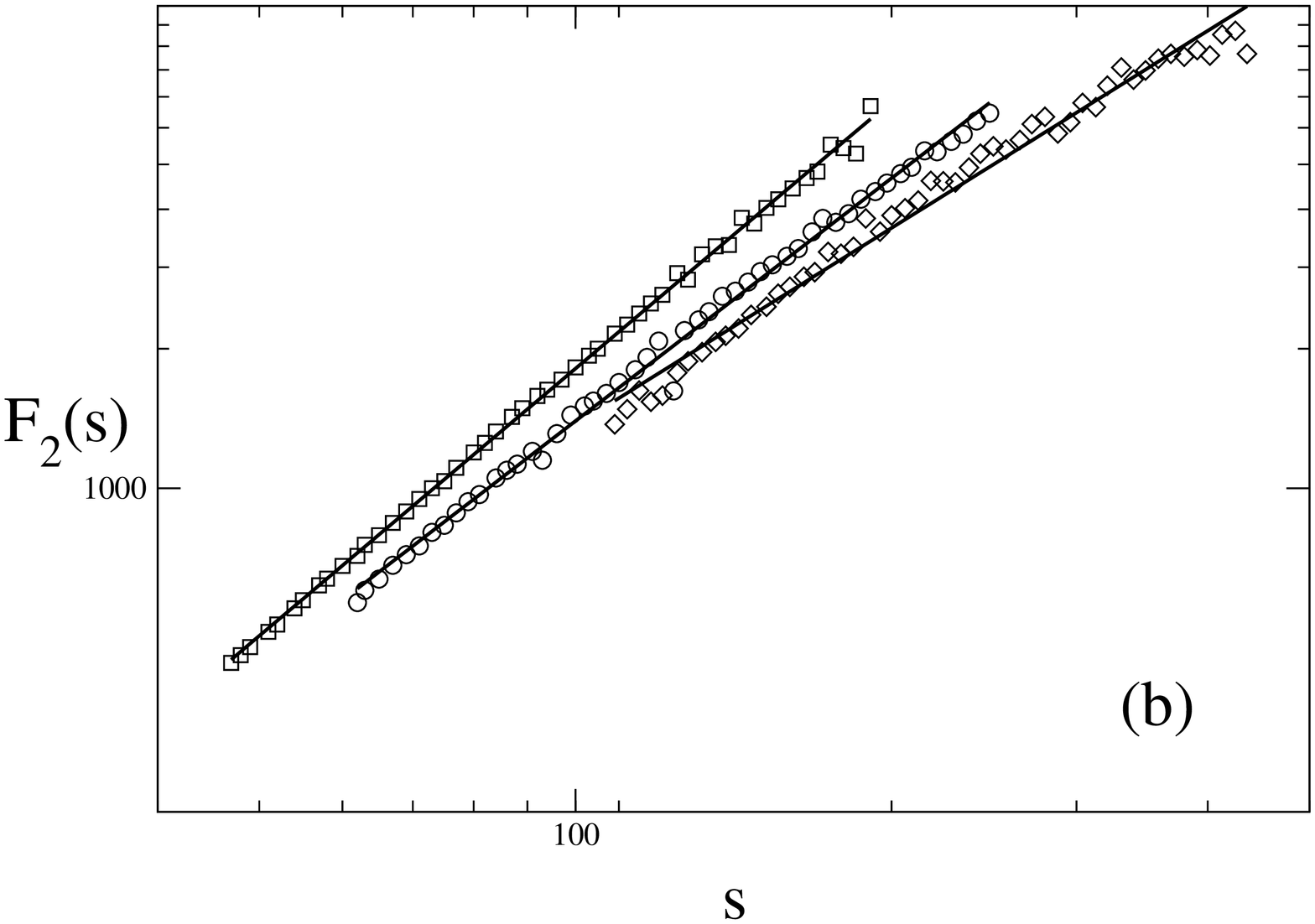}}
\caption{ Panel (a): Typical forms of the fluctuation function for intermaxima time series with
and without scaling properties. Fluctuation function $F_{2}(s)$ for
the parents are as follows. Circles: female parent. Squares: male parent.  For these time series
r.m.s. fit of the power law is shown as a continuous straight line.
The closeness to a straight line on the log-log scale means that the
corresponding time series of the intermaxima intervals  have scaling properties.
For comparison typical time series for a healtly Drosophila (filled triangles) is presented.
We do not observe scaling and thus we cannot calculate any fractal spectra.
Panel (b): Fluctuation functions $F_{2}(s)$ and power-law r.m.s.
fits (solid lines) for time series of the first generation of flies
(the kids).  As we see there is no drastic breaking of the scaling as it is for the
healtly Drosophila of panel (a).
}\label{fig:03}
\end{figure}
\par
Panel (a) of Fig. 3 shows the fluctuation functions ( $q=2$) for a
healtly Drosophila and for parent flies with heart defects. In the case of
humans the normal sinus rhythm of the heartbeat activity has
complex behavior similar to the behavior of a chaotic attractor
\cite{bassin}. The heart dynamics of humans with heart diseases
may become more periodic in comparison to the heartbeat dynamics
of the healtly individuals. The heartbeat dynamics of the
investigated here Drosophila shows opposite behavior. We  see that
the fluctuation function for the healtly Drosophila does not exhibit
scaling at least for small $s$ and this lack of scaling is
observed for all values of the parameter $q$. The deviation from
the scaling behavior for the fluctuation function means that we
can not calculate any fractal spectra for the healtly Drosophila
opposite to the case of  the flies with genetic defects where
the fluctuation function can show good scaling properties for
the whole studied range of $s$. We note that the fluctuation functions for the
parents seem to be very close to a straight line on a log-log scale.
Thus we shall proceed with calculation of the fractal spectra.
These spectra will have different properties for time series of Drosophila with
different heart defects.
\par
\begin{figure}[t]
\centerline{\includegraphics[angle=270,width=12cm]{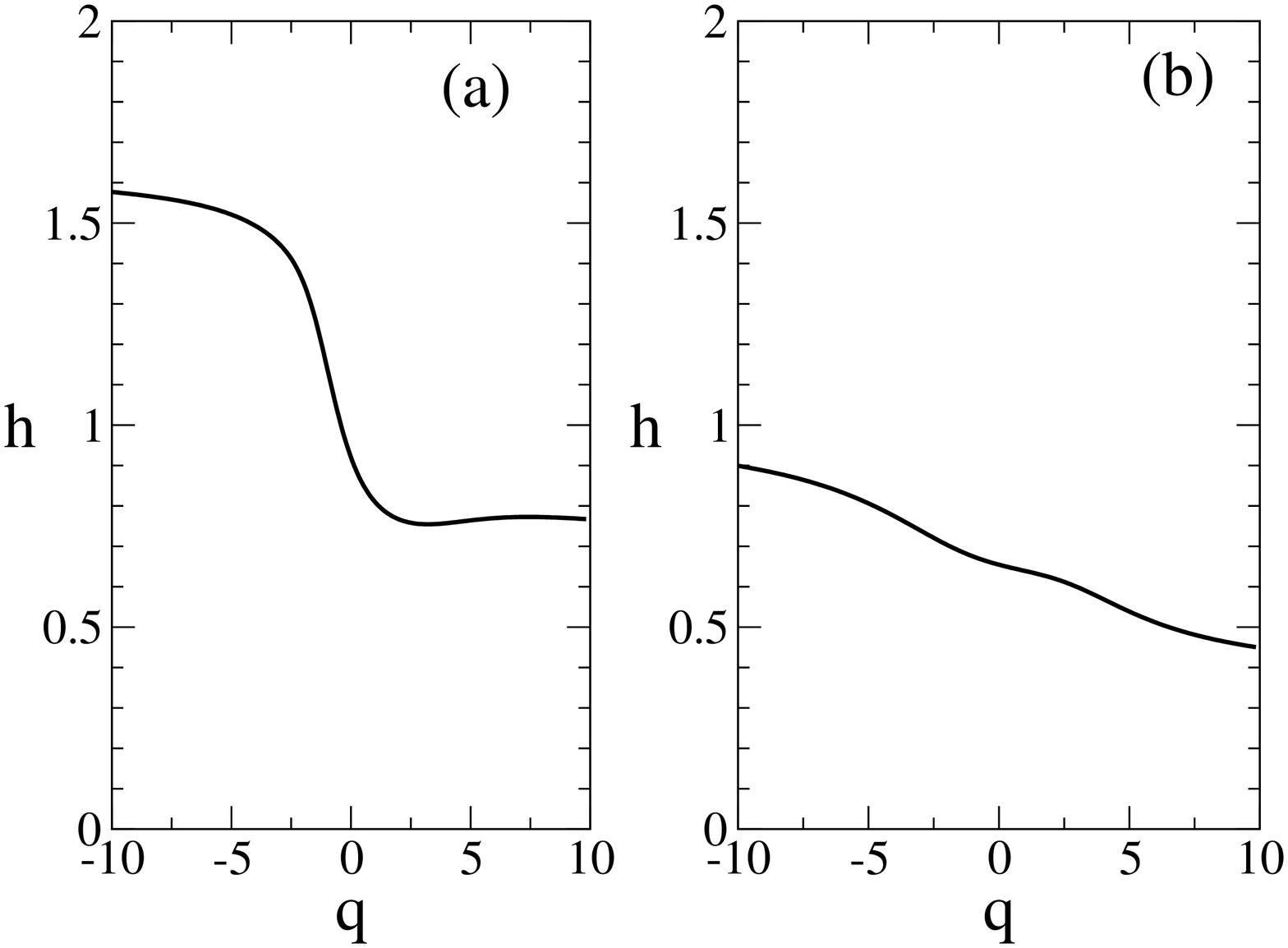}}
\caption{ The fractal spectrum $h(q)$ for the time series of the parents.
$q$ changes continuously from $-10$ to $10$. Panel (a): female parent. Panel (b):
male parent
}\label{fig:04}
\end{figure}
The Hurst exponent for the two parents is presented in Fig. 4.
$h$ is not a constant and hence the two time series of the parents have multifractal
properties. Thus multifractal cardiac dynamics can be observed not
only for humans but also for much simpler animals like Drososphila.
\begin{figure}[t]
\centerline{\includegraphics[angle=270,width=12cm]{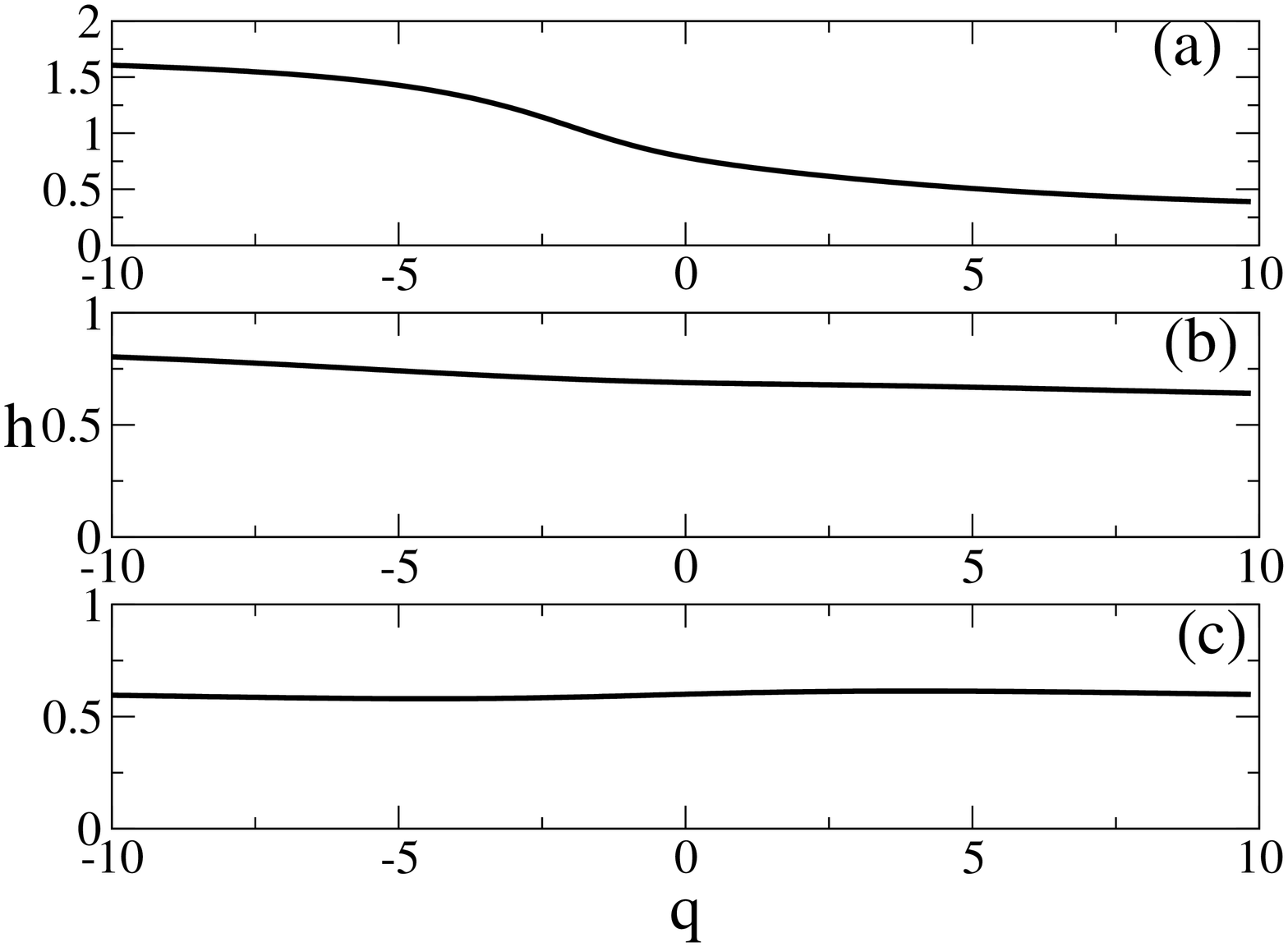}}
\caption{ The fractal spectrum $h(q)$ for the time series of the first generation
(the kids). From the top to the bottom the three characteristic shapes of this spectra
are shown.
}\label{fig:05}
\end{figure}
In  figures 5 and 6 we see the kinds of spectra of
the Hurst exponent characteristic for the first and second
generations of flies obtained from the above-mentioned parents with
genetically defect hearts. The  spectra in panels (a) and (b) in
Fig. 5 are of the same kinds as the spectra of the two parents.
The spectrum in panel (c) has nontypical from the point of view of
physics because in  most physical systems $h$ decreases
with increasing $q$.
\begin{figure}[t]
\centerline{\includegraphics[angle=270,width=10cm]{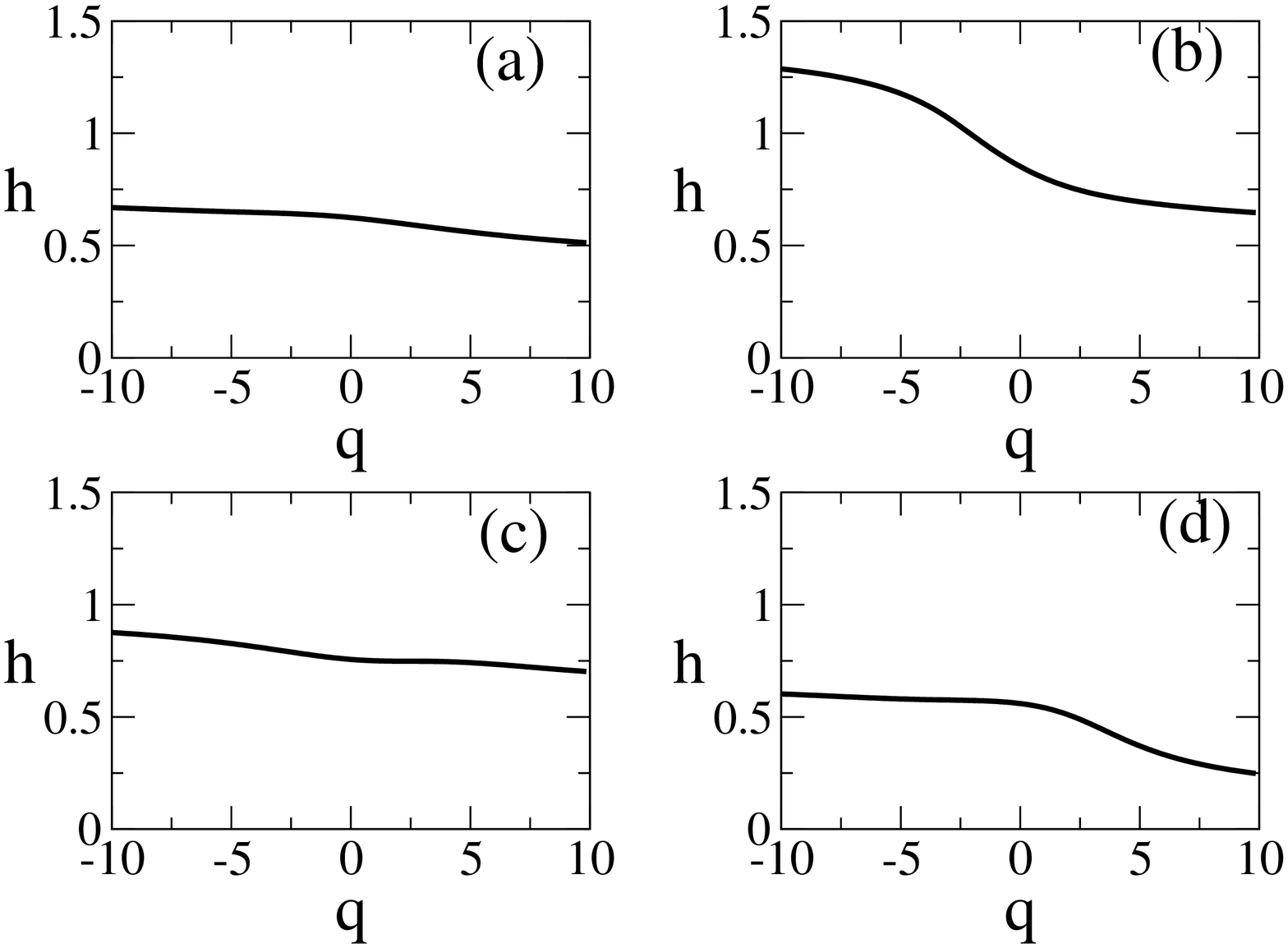}}
\caption{ The fractal spectra $h(q)$ for the time series of the second generation
(the kids of the kids).
}\label{fig:06}
\end{figure}
For the second generation of flies we observe the two kinds of
$h(q)$ spectra existing in the case of the parents plus an
additional kind of spectrum with Hurst exponent which is
systematically smaller than $0.5$ for positive $q$ i.e. the
anticorrelations  dominate the corresponding time series.
\begin{figure}[t]
\centerline{\includegraphics[angle=270,width=10cm]{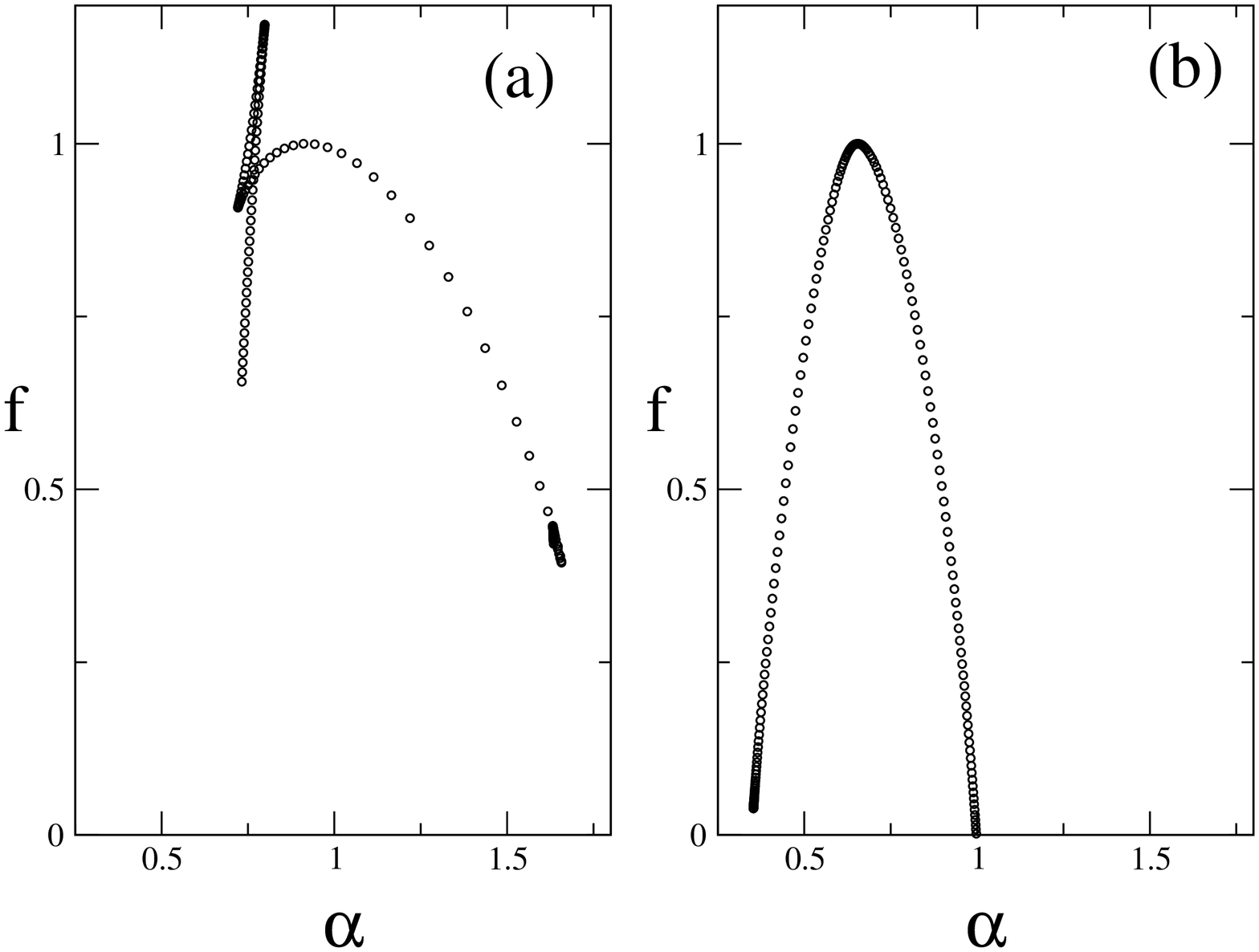}}
\caption{ The fractal spectrum $f(\alpha)$ for the time series of
the parents. Panel (a): female parent; Panel (b): male parent.
Parameters of the spectrum in panel (b) are: $\alpha_{min}=0.35$,
$\alpha_{max}=1.01$, $f_{max}=1.00$ at $\alpha=0.66$.
$\alpha_{l}(0.9 f_{max})=0.57$, $\alpha_{r}(0.9 f_{max})=0.75$.
Thus the width $\Delta(0.9 f_{max})=0.18$. }\label{fig:07}
\end{figure}
\par
The difference in the dynamical properties of the  intermaxima time series for the
heartbeat activity of Drosophila  can be investigated by means of their $f(\alpha)$ spectra.
Fig. 7 shows these spectra for the parents. For the spectra
with parabolic form, the parts of elements of the
time series with a given value of $\alpha$, build a partial fractal with a fractal
dimension denoted by $f(\alpha)$. The top part of the spectrum
which is located around some
value $f(\alpha^{*})$ corresponds to the statistical most significant part of the spectrum
(corresponding to the parts of the time series with the largest dimension). $f(\alpha^{*})$
gives the value of this largest dimension and we can distinguish the time series with
respect to the value of $\alpha^{*}$ and the width of the spectrum around the maximum (
$\Delta= \alpha_{r}(f^{*}) - \alpha_{l}(f^{*})$, where $f^{*}$ is characteristic
which we shall take to be equal of $0.9 f_{max}$ in order to compare the
parameters of the $f(\alpha)$ spectra of all generations of Drosophila.
$\alpha_{l}$ and $\alpha_{r}$ are the values of $\alpha$ corresponding
to $f^{*}$ and positioned to the left and to the right with respect to the
value $\alpha^{*}$ corresponding to the maximum of the
$f(\alpha)$ spectrum).
Wide $f(\alpha)$ spectrum corresponds to more distributed multifractal (the
partial fractal dimensions are less concentrated around the maximum
partial dimension $f_{max}$) and a
narrow spectrum corresponds to more concentrated multifractal. Coming back to the spectra of
parents in Fig. 7 we observe the typical parabolic form of the spectrum only for the
male parent. Thus the form of the $f(\alpha)$ spectrum can help us to distinguish
among the heart defects of Drosophila as some of these defects
 (and in particular the genetic defect of the female
parent) can lead to nonparabolic form of the $f(\alpha)$ spectrum, i.e.,
to deviation from the ideal multifractal behaviour.
 $f_{max}=1$ for the spectrum
of the male parent and its $0.9 f_{max}$ width is $\Delta=0.18$.
The result of the combination of
the two kinds of dynamics leading to parabolic and nonparabolic $f(\alpha)$ spectra can be observed in the
spectra of the two generation of flies following the parents. The characteristic spectra
for the second generation are presented in Fig. 8. We observe two kinds of consequences
from the form of the spectrum of the female parent (i)
the nonparabolic kind of spectrum is reproduced as it can be
seen in panel (b) of Fig. 8. and (ii) some (but not al) of the parameters of the
parabolic spectra change.
We note that for the parabolic spectra in panels (a), (c), (d) of Fig. 8
$f_{max}$ remains unchanged and equal to $1$ not only  for this
generation of flies but also for the parabolic spectra in the next
generation shown in Fig. 9.  For the second generation of flies
$\alpha$ for $f_{max}$ is dispersed around $0.66$ -
its value for the male parent.

\begin{figure}[t]
\centerline{\includegraphics[angle=270,width=12cm]{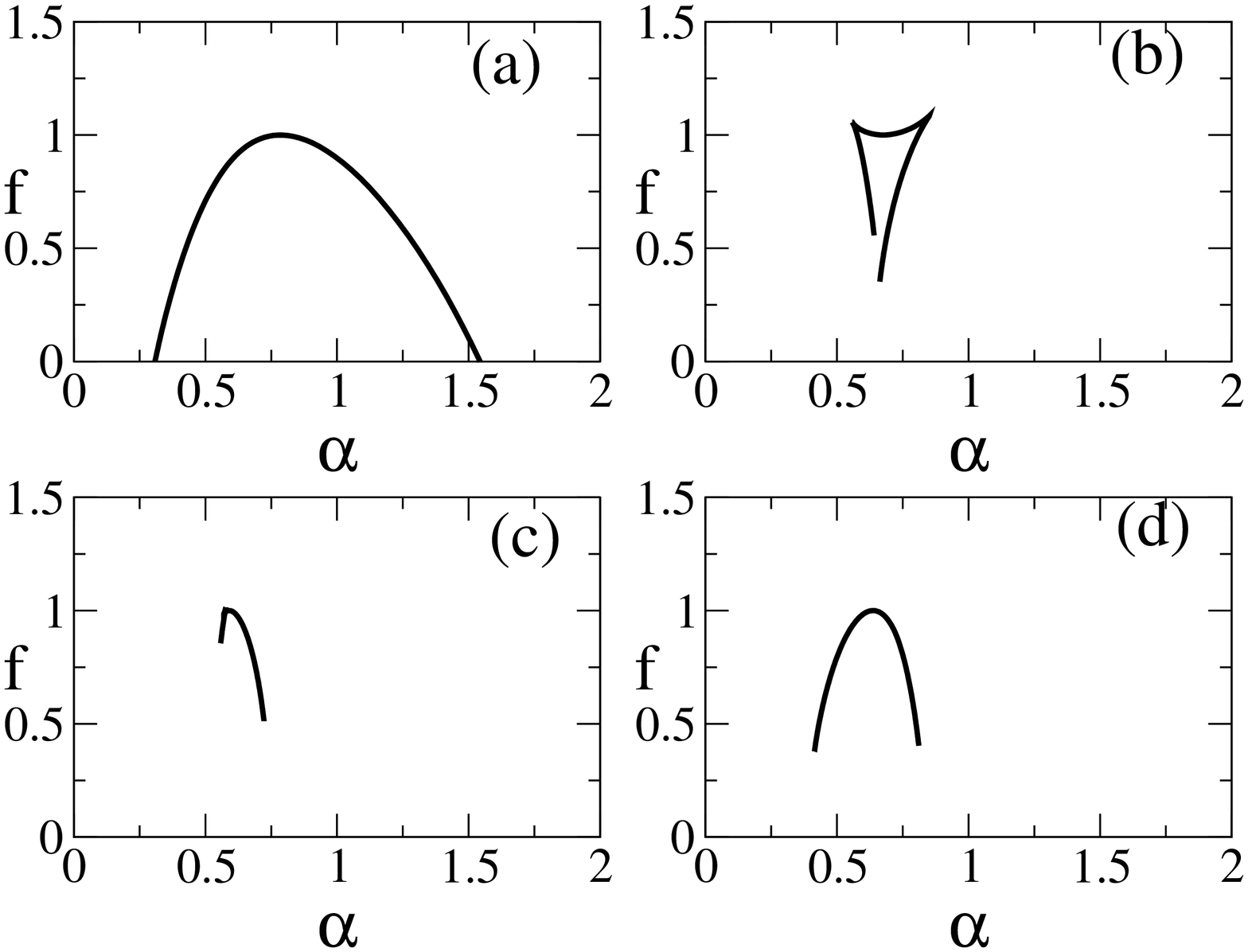}}
\caption{ The fractal spectra $f(\alpha)$ for the time series of the second generation
(the kids).  Four characteristic shapes of this spectra are shown. The parameters of the
spectra of parabolic kinds are: Panel (a): $\alpha_{min}=0.24$, $\alpha_{max}=1.79$,
$\alpha_{l}=0.61$, $\alpha_{r}=0.99$, $\Delta=0.28$. $f_{max}=1.00$ at $\alpha=0.80$.
Panel (c): $\alpha_{min}=0.56$, $\alpha_{max}=0.72$, $\alpha_{l}=0.56$,
$\alpha_{r}=0.65$, $\Delta=0.08$ $f_{max}=1.00$ at $\alpha=0.59$. Panel (d):
$\alpha_{min}=0.41$, $\alpha_{max}=0.81$, $\alpha_{l}=0.54$, $\alpha_{r}=0.72$,
$\Delta=0.18$. $f_{max}=1.00$ at $\alpha=0.65$.
}\label{fig:08}
\end{figure}
\begin{figure}[t]
\centerline{\includegraphics[angle=270,width=12cm]{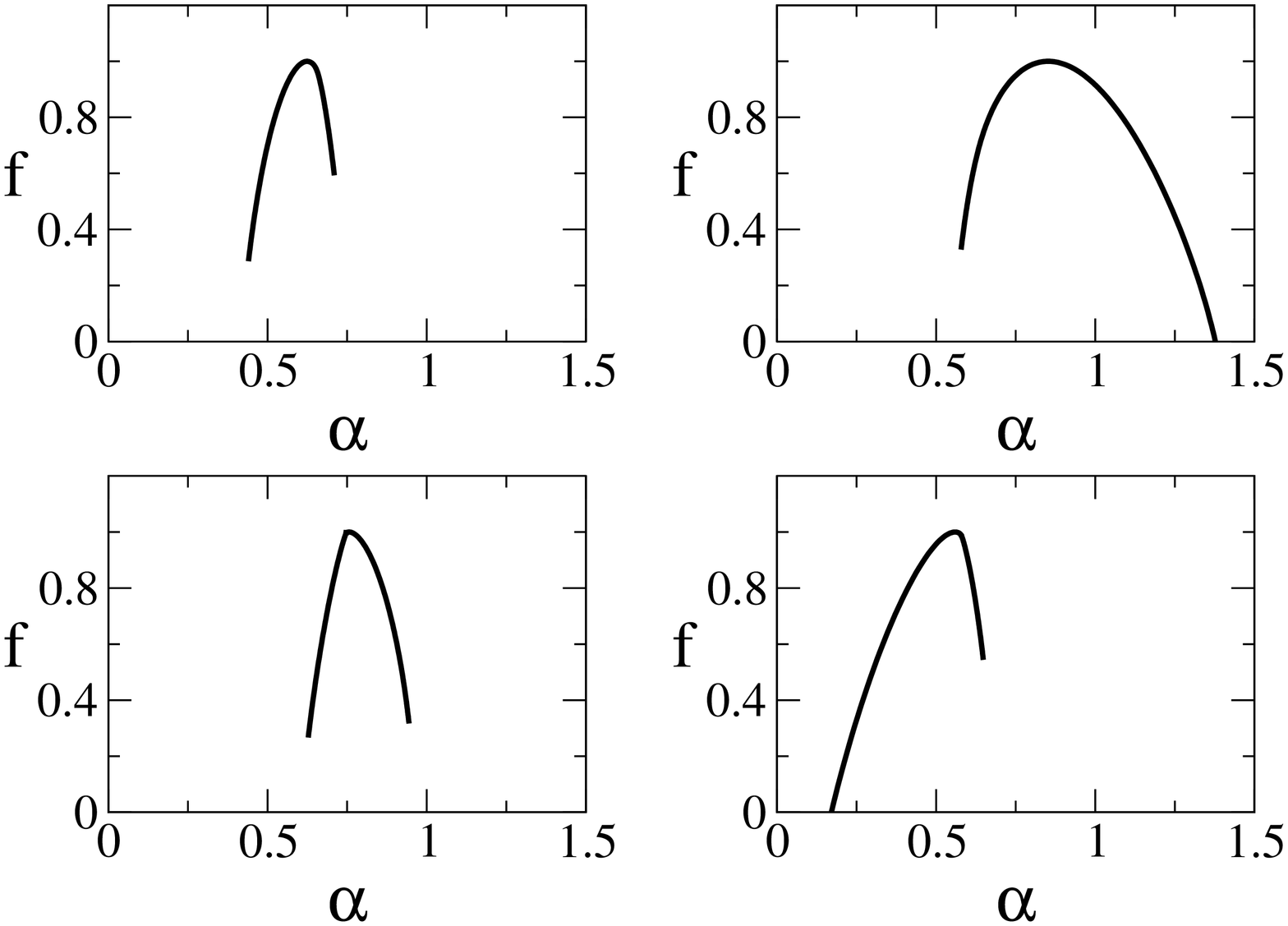}}
\caption{ Parabolic fractal spectra $f(\alpha)$ for the time series of the third generation
(the kids of the kids).  Four characteristic shapes of this spectra are shown. Parameters
of the spectra are as follows: Panel (a): $\alpha_{min}=0.44$, $\alpha_{max}=0.71$,
$\alpha_{l}=0.55$, $\alpha_{r}=0.67$, $\Delta=0.12$. $f_{max}=1.00$ at $\alpha=0.65$. Panel (b): $\alpha_{min}=0.58$, $\alpha_{max}=1.40$,
$\alpha_{l}=0.72$, $\alpha_{r}=1.01$, $\Delta=0.29$. $f_{max}=1.00$ at $\alpha=0.86$. Panel (c): $\alpha_{min}=0.63$, $\alpha_{max}=0.94$,
$\alpha=0.72$, $\alpha_{r}=0.83$, $\Delta=0.13$.$f_{max}=1.00$ at $\alpha=0.78$. Panel (d): $\alpha_{min}=0.12$, $\alpha_{max}=0.65$,
$\alpha_{l}=0.47$, $\alpha_{r}=0.60$, $\Delta=0.13$. $f_{max}=1.00$ at $\alpha=0.56$.
}\label{fig:09}
\end{figure}
In the third generation of flies the nonparabolic form of the spectrum is
reproduced again. From several characteristic
examples of parabolic spectra of this generation which are shown in Fig.9
only one of the spectra has a wide basis. For all spectra $f_{max}=1$ and
for the spectra from panels (a), (c), (d) $\Delta$ is almost the same.
\section{Concluding remarks}
In this paper we apply the multifractal detrended fluctuation
analysis (MFDFA)  to the study of  Drosophila ECG time series.
On the example of Drosophila we have shown that
the presence of long-range correlations in the heartbeat activity is property
not only of humans and complex animals and can be observed in much simpler
animals as for example in Drosophila melanogaster. Opposite to
the heartbeat dynamics of healtly humans which is
described by broad range of Hurst exponents the intermaxima
intervals of the time series of the heartbeat dynamics of healtly
Drosophila do not have scaling properties and thus it cannot be
described by means of scaling exponents and fractal spectra. We
have shown that the presence of genetic defects can lead to
long-range correlations of the heartbeat dynamics of Drosophila.
The transfer of the multifractal properties from generation to
generation and the similarity of the kinds and parameters of the
multifractal spectra for different generations of Drosophila show
that a correlation could exists between genetic properties and
dynamic patterns in the heartbeat activity of simple animals like
Drosophila. We can conjecture that the above correlation exists for
the case of other simple animals and probably also for the case of
more complex animals and ever humans.
\section*{Acknowledgements}
N. K. V. gratefully acknowledges the support by the Alexander von
Humboldt Foundation and by NSF of Republic of Bulgaria (contract MM 1201/02).
E.D.Y. thanks the  EC Marie Curie Fellowship Programm
(contract  QLK5-CT-2000-51155) for the support of her research.


\begin{thebibliography}{}
\small
\bibitem{bassin}
Bassinngthwaighte J. B., Liebovitch L. S., West B. J. (1994). Fractal physiology.
(Oxford University Press: New York).

\bibitem{ivanov1}
Ivanov P. Ch. (2003). Long-range dependence in heartbeat dynamics, p.p. 339-368
in Ragarajan G. and Ding H (eds.) (2003). Processes with long-range correlations.
Lecture Notes in Physics , vol. 621 (Springer: Berlin).

\bibitem{skinner1}
Skinner, J.E., Pratt, C.M., and Vybiral, T.A.  (1993).
A reduction in the correlation dimension of heart  beat  intervals  proceeds
imminent ventricular  fibrillation  in human subjects.
American Heart Journal {\bf 125}, 731-743.

\bibitem{johnson}
Jonson, E., Ringo, J. and Dowse, H. (2001) Dynamin, encoded by shibire, is central to cardiac function.
Journal of Experimental Zoology, {\bf 289},  81-89.

\bibitem{mandelbrot}
Mandelbrot B. B. (1982). The fractal geometry of the Nature. (Freeman: San Francisco).


\bibitem{stanley1}
Stanley H. E. (1999). Scaling, universality, and renormalization: Three pilars of the modern critical
phenomena. Rev. Mod. Phys. {\bf 71} S358-S366.

\bibitem{stanley2}
Stanley H. E., Buldyrev S. V., Goldberger A. L., Goldberger Z. D., Havlin S.,
Mantegna R. N., Ossadnik S. M., Peng C.-K., Simons M. (1994). Statistical mechanics
in biology: how ubiquitous are long-range correlations. Physica A {\bf 205}, 214-253.

\bibitem{tel}
Tel T. (1988). Fractals, multifractals and thermodynamics. Zeitschrift f{\"u}r
Naturforschung A {\bf 43}, 1154-1174.

\bibitem{peitgen}
Everetsz C. J. G., Mandelbrot B. B. (1992). Multifractal measures. p.p. 921-953
in Peitgen H. -O., J\"{u}rgens, Saupe D. Chaos and fractals. New frontiers of
science. Springer, New York.

\bibitem{ott}
Ott E. (1993).
Chaos in dynamical systems. (Cambridge University Press: Cambridge).

\bibitem{muzy}
Muzy J. F., Bacry E., Arneodo A. (1993). Multifractal formalism for fractal signals.
The structure function approach versus the wavelet-transform modulus-maxima method.
Physical Review E {\bf 47}, 875-884.

\bibitem{arneodo}
Muzy J. F., Bacry E., Arneodo A. (1994).
The multifractal formalism, revisited with wavelets. Interantional Journal of Bifurcations
and Chaos. {\bf 4}, 254-302.


\bibitem{arneodo96}
Arneodo A., d'Aubenton-Garafa Y., Graves P. V., Muzy J. F., Thermes C. (1996)
Wavelet based fractal analysis of DNA sequences. Physica D {\bf 96}, 291-320.

\bibitem{arneodo99}
Arneodo A., Manneville S., Muzy J. F., Roux S. G. (1999).
Revealing a lognormal cascading process in turbulent velocity statistics with
wavelet analysis. Philosophical Transactions of the Royal Society of London A
{\bf 357}, 2415-2438.

\bibitem{arneodo03}
Arneodo A., Decoster N., Kestener P., Roux S. G. (2003).
A Wavelet-based method for multifractal image analysis: From theoretical concepts
to experimental applications. Advances in Imaging and Electron Physics
{\bf 126}, 1-92.


\bibitem{dimitrova}
Dimitrova Z. I., Vitanov N. K. (2004).
Chaotic pairwise competition. Theoretical Population Biology. {\bf 66}, 1-12.

\bibitem{kantelhardt}
Kantelhardt J. W., Zschiegner S. A., Koscielny-Bunde E., Havlin S., Bunde A.,
Stanley H. E. (2002). Multifractal detrended fluctuation analysis of nonstationary
time series. Physica A {\bf 316}, 87-114.

\bibitem{peng1}
Peng C. -K., Mietus J., Hausdorf J. M., Havlin S.,Stanley H. E., Goldberger A. L.
(1993). Long-range anticorrelations and non-Gaussian behavior of the heartbeat.
Phys. Rev. Lett. {\bf 70}, 1343-1346.


\bibitem{ivanov2}
Ivanov P. Ch., Amaral L. A. N., Goldberger A. L., Havlin S., Rosenblum M. G.,
Struzik Z. R., Stanley H. E. (1999). Multifractality in human heartbeat dynamics.
Nature {\bf 399}, 461-465.

\end{thebibliography}
\end{document}